\documentclass[prx,aps,twocolumn,showpacs,amsmath,amssymb]{revtex4-1}
\usepackage{times}
\usepackage{textcomp}
\usepackage{graphicx}
\usepackage{bm}
\allowdisplaybreaks 
\usepackage{braket}
\usepackage[breaklinks]{hyperref}
\hypersetup{colorlinks=true, linkcolor=blue, citecolor=blue, filecolor=blue, urlcolor=blue}

\begin{document}

\title{Bulk spectroscopic measurement of the topological charge of Weyl nodes with resonant x-rays}

\author{Stefanos Kourtis}
\affiliation{Department of Physics, Princeton University, Princeton, NJ 08544, USA}

\date{\today}

\begin{abstract}
We formulate a bulk spectroscopic method for direct measurement of the topological invariant of nondegenerate band crossings in Weyl semimetals. We demonstrate how polarization-resolved resonant x-ray scattering captures the winding of the Berry flux around Weyl nodes. The spectra obtained by the proposed strategy feature an integer number of zero-crossings that is directly related to the topological charge of the measured nodes. We benchmark the proposed protocol on TaAs, using realistic low-energy models derived from density-functional theory to accurately represent the states close to the Fermi level, including sizable deviations from the idealized linear dispersion. We conclude that the proposed measurement, which is within the reach of current experimental setups, yields direct signatures of nontrivial band topology in spectroscopy of three-dimensional bulk matter.
\end{abstract}

\maketitle

\section{Introduction}

The two defining hallmarks of topological insulators are their distinctive bulk transport responses and the presence of current-carrying states confined at their boundaries~\cite{Hasan2010,Qi2008}. The latter feature has rendered surface spectroscopy a well-established method for characterization of three-dimensional topological states of matter. On the other hand, spectral quantities in the bulk of a topological insulator are generally expected to be indistinguishable from those of a trivial insulator. However, topological semimetals feature a finite number of topologically nontrivial Fermi surfaces in the bulk~\cite{volovikbook,Turner2013,Hasan2014}. Weyl semimetals (WSMs), in particular, possess pairs of nondegenerate linear band crossings at isolated points in the Brillouin zone (BZ) called Weyl nodes. There, electronic properties are effectively governed by the relativistic Weyl equation. Functioning as sources and sinks of Berry flux, these points give rise to a number of remarkable physical properties, including open constant-energy contours in the surface band structure called Fermi arcs~\cite{Wan2011,Xu2011,Balents2011,Bernevig2015} and the condensed-matter realization of the chiral anomaly in the bulk~\cite{Nielsen1983,Aji2012,Zyuzin2012a,Hosur2013a,Kim2013,Parameswaran2014,Behrends2015}.

Confirming theoretical predictions~\cite{Weng2015,Huang2015c}, angle-resolved photoemission spectroscopy (ARPES) has identified Ta and Nb monarsenides and monophosphides as the first realizations of WSMs in solids~\cite{Lv2015a,Lv2015b,Yang2015,Xu2015a,Xu2015b,Belopolski2016}. More recent experiments~\cite{Belopolski2015a,Deng2016,Xu2016,Jiang2016,Liang2016,Xu2016a} have unearthed representatives of a subsequently identified class of WSMs labeled ``type-II''~\cite{Soluyanov2015,Sun2015,Wang2015}. Particular emphasis has been placed on resolving spectral signatures of topological origin at the boundary: ARPES has been used to identify surface bands as Fermi arcs~\cite{Belopolski2016,Lv2015a,Lv2015b,Yang2015,Xu2015a,Xu2015b,Belopolski2016,Deng2016,Xu2016,Jiang2016} and scanning tunneling spectroscopy (STS) has been employed to search for uniquely characterizing patterns~\cite{Kourtis2015a} in surface quasiparticle interference (QPI)~\cite{Zheng2015,Inoue2016,Batabyal2016,Deng2016}. Magnetotransport measurements~\cite{Huang2015,Zhang2015,Zhang2015a,Klotz2015,Arnold2015,Zhang2016,Arnold2016}, on the other hand, have revealed a negative magnetoresistance and quantum oscillations that are compatible with the presence of Weyl nodes in the bulk.

Despite the successful identification of WSM states in more than one material classes, the topological content of individual Weyl nodes is still inaccessible to experiment directly. Quantum oscillations~\cite{Murakawa2013}, as well as recently proposed transport methods~\cite{Dai2015}, have the capacity to access the Berry phase of Fermi surfaces surrounding Weyl nodes, whenever the chemical potential is favorably placed with respect to the nodal point. However, contributions from similarly sized Fermi surfaces are superimposed in these methods and the overall response comes from the entire BZ. Moreover, these approaches yield no response at all when the chemical potential lies at the nodal point. Experiments have so far inferred the topological charge from the number and dispersion of edge modes originating from the projections of Weyl nodes at the boundary~\cite{Lv2015a,Yang2015,Xu2015a,Xu2015b}. This approach is not always viable~\cite{Belopolski2016} and is impeded by overlaps of trivial boundary Fermi surfaces, projections of bulk Fermi surfaces and Fermi arcs on a given surface. Additionally, it may not be possible to cleave the surface of interest cleanly enough to observe the Fermi arcs with ARPES.

The purpose of this work is to show that the topological nature of a WSM can be revealed by a targeted spectroscopic measurement of Weyl nodes in the bulk using polarization-resolved resonant inelastic x-ray scattering (RIXS)~\cite{Kotani2001,Ament2011}. The unique spin-orbital selectivity of polarized RIXS allows for the measurement of the effective spinor components close to a Weyl point. The topological charge of a Weyl node manifests itself as zero-crossings in suitable combinations of RIXS spectra, obtained using different photon beam polarizations. The number of these zero-crossings is found to be a fixed integer times the topological charge of the targeted nodes. This scheme is robust against sizable perturbations away from the idealized linearly dispersive regime. The rest of this paper is organized as follows: In Sec.~\ref{sec:theory} we introduce the low-energy RIXS cross section in Weyl semimetals. In Sec.~\ref{sec:meas} we provide the theoretical description for a direct bulk measurement of the topological invariant characterizing a Weyl node using RIXS. We then study the robustness of the pertinent spectral signature under arbitrary deformations of the scattering geometry and perturbations away from the linear energy dispersion in Sec.~\ref{sec:geom}. In Sec.~\ref{sec:taas} we apply the measurement protocol to the Weyl nodes in TaAs and show that it indeed yields the topological invariant of the targeted nodes in a material-specific application. Finally, we discuss in detail the feasibility of the experimental procedure in Sec.~\ref{sec:exp}.

\section{Low-energy RIXS response in Weyl semimetals}\label{sec:theory}

\subsection{Low-energy theory around Weyl nodes}

In WSMs, the interesting physics happens close to the nodal points of the band structure. There, only two bands are relevant and the effective hamiltonian is
\begin{equation}
{\cal H} = \sum_{\bm{k}\in\mathrm{BZ}} \psi_{\bm{k}}^\dagger H_{\bm{k}} \psi_{\bm{k}} \,,
\end{equation}
where $\psi_{\bm{k}} = (c_{\bm{k},\uparrow} \ c_{\bm{k},\downarrow} )^\mathsf{T}$ is a spinor containing annihilation operators acting on electrons with wavevector $\bm{k}$ and pseudospin $\sigma=\uparrow,\downarrow$. The $2\times2$ hamiltonian matrix is
\begin{equation}
H_{{\bm{k}}} = \bm{g}_{\bm{k}} \cdot \bm{\tau} + (g_{0,\bm{k}} + \mu) \tau_{0} \,,\label{eq:2band}
\end{equation}
where the $2\times2$ identity matrix $\tau_{0}$ and the three Pauli matrices $\boldsymbol\tau=(\tau_{x},\tau_{y},\tau_{z})$ span the pseudospin basis, $\bm{g}_{\bm{k}} = ( g_{1,\bm{k}},g_{2,\bm{k}},g_{3,\bm{k}} )$ and $\mu$ an overall chemical potential. $g_{0,\bm{k}}$, $g_{1,\bm{k}}$, $g_{2,\bm{k}}$, and $g_{3,\bm{k}}$ are real-valued functions of the wave-number $\bm{k}$. The spectrum is $\varepsilon_{\bm{k},\pm} = g_{0,\bm{k}} + \mu \pm g_{\bm{k}}$ with $g_{\bm{k}}=|\bm{g}_{\bm{k}}|$. In what follows, the pseudospin will arise from the \textit{orbital} degree of freedom of electrons.

The topological invariant pertinent to Weyl nodes can be defined as~\cite{volovikbook}
\begin{equation}
 C = \frac{1}{2\pi} \oint_\Omega \mathrm{d}\bm{k} \, \frac{\bm{g}_{\bm{k}}}{2g_{\bm{k}}^3} \cdot ( \partial_{\lambda} \bm{g}_{\bm{k}} \times \partial_{\nu} \bm{g}_{\bm{k}} ) \,,\label{eq:chern}
\end{equation}
where $\Omega$ denotes a closed two-dimensional surface that contains the Weyl node and $\lambda, \nu$ are coordinates that parametrize the surface. The integrand is the Berry flux through $\Omega$~\cite{Xiao2010}.
A Weyl node carrying a topological charge $C=1$, henceforth called a single Weyl node, is described by
\begin{equation}
\bm{g}_{\bm{k}}^{\mathrm{single}} = (v_x k_x, v_y k_y, v_z k_z) \,,\label{eq:1weyl}
\end{equation}
where $v_i$, $i=x,y,z$ are momentum-space velocities. For Weyl nodes with $C=2$, or double Weyl nodes~\cite{Onoda2002},
\begin{equation}
 \bm{g}_{\bm{k}}^{\mathrm{double}} = ( v_x^2 k_x^2 - v_y^2 k_y^2, 2 v_x v_y k_x k_y, v_z k_z ) \,.\label{eq:2weyl}
\end{equation}
Eq.~\eqref{eq:chern} implies the winding of the Berry flux around the Weyl node. When $|C|=1$, the Berry flux winds once along any path ${\cal P}$ on $\Omega$ that encircles the Weyl point. For $\bm{g}_{\bm{k}}^{\mathrm{double}}$ the winding is observed when the projection of ${\cal P}$ in the $k_z = 0$ plane encloses the Weyl node. The Berry flux then winds twice along ${\cal P}$~\cite{Xu2011,Fang2012a}. Since $\bm{g}_{\bm{k}}$ is aligned to the Berry flux, a measurement of the former yields the latter. We will now show how the components of $\bm{g}_{\bm{k}}$ can be identified in RIXS.

\subsection{RIXS cross section}

In a RIXS experiment, core electrons of a specific ion are promoted to an unoccupied state using an intense x-ray beam, thereby locally exciting the irradiated material into a highly energetic and very short-lived intermediate state~\cite{Kotani2001,Ament2011}. The core and valence spin-orbital states that participate in the scattering can be selected by the choice of incoming and outgoing photon polarizations. Subsequently, the core hole recombines with a valence electron, after a lifetime of the order of 1 femtosecond. The process imparts both energy and momentum to particle-hole excitations left behind in the material, as the momentum and energy of decaying electrons are generally not the same as those of the photoexcited one. Their dispersion can be inferred by the differences in scattering angle and frequency of incoming and outgoing x-ray photons.

The total RIXS intensity at zero temperature is~\cite{Kramers1925,Blume1985,Ament2011,Ament2011a}
\begin{align}
 I(\bm{q},\omega_{\bm{k}},\omega_{\bm{k}'},\boldsymbol\epsilon,\boldsymbol\epsilon') = \sum_{fg} & |{\cal F}_{fg}(\boldsymbol\epsilon,\boldsymbol\epsilon',\bm{q},\omega_{\bm{k}})|^2 \nonumber\\
 & \times \delta(E_g - E_f + \hbar\omega_{\bm{kk}'}) \,,\label{eq:cross}
\end{align}
where $\hbar\omega_{\bm{kk}'}=\hbar(\omega_{\bm{k}'}-\omega_{\bm{k}})$ and $\bm{q}=\bm{k}'-\bm{k}$ are the energy and momentum transferred to the material, $\bm{k}$ and $\bm{k}'$ ($\boldsymbol\epsilon$ and $\boldsymbol\epsilon'$) the incoming and outgoing photon momenta (polarizations) and $E_g$ and $E_f$ the eigenenergies corresponding to initial and final states $\ket{g}$ and $\ket{f}$. The scattering amplitude ${\cal F}_{fg}$ contains dipole operators that describe the transitions between core and unoccupied levels. Here we consider the processes in which core electrons are promoted directly into the orbitals that generate the physics of interest. This experimental setting is frequently referred to as ``direct'' RIXS. In the monarsenide / monophosphide family of WSMs, the bands at the Weyl nodes are contributed predominantly by Ta or Nb ions. We therefore consider the $L$ or $M$ edges of Nb or Ta, which correspond to transitions $2p / 3p \rightarrow 4d / 5d$.

The intermediate state in RIXS is typically very short-lived. Due to this, the fast-collision approximation can be employed to accurately model the scattering process~\cite{Hannon1988,Luo1993,vanVeenendaal2006}. In the fast-collision approximation, the inelastic scattering amplitude at a specific absorption edge is defined as~\cite{Haverkort2010b,Haverkort2010,Ament2011,Marra2012}
\begin{align}
 {\cal F}_{fg}(\boldsymbol\epsilon,\boldsymbol\epsilon',\bm{q},\omega_{\bm{k}}) =&{\ } \bra{f} \sum_{\bm{k},s,s'} c_{\bm{k},s'} T_{s',s}(\boldsymbol\epsilon,\boldsymbol\epsilon',\omega_{\bm{k}}) c_{\bm{k}+\bm{q},s}^\dagger \ket{g} \,.\label{eq:rixsdef}
\end{align}
where $s$ and $s'$ are combined indices of spin and orbital degrees of freedom and the operator $c_{\bm{k},s}^\dagger (c_{\bm{k},s})$ creates (annihilates) an electron in the single-particle state $\ket{\bm{k},s}$, and $\boldsymbol\epsilon$ and $\boldsymbol\epsilon'$ are the incoming and outgoing beam polarizations, respectively. The entries of the complex matrix $T$ contain the fundamental absorption cross sections for spin-preserving and spin-flip processes, which also depend on the core orbitals~\cite{Haverkort2010b,Haverkort2010}. Note that core-hole dynamics is insignificant here, so the operators acting on core electrons have been dropped.

An appropriate choice of polarizations can isolate the components of the (pseudo)spin density. As a simple example, consider the measurement of spin winding in a WSM with tetragonal symmetry. To first order, the overall polarization dependence is given by the matrix~\cite{Haverkort2010b}
\begin{subequations}
\begin{equation}
 T(\boldsymbol\epsilon,\boldsymbol\epsilon',\omega_{\bm{k}}) = T_0(\boldsymbol\epsilon,\boldsymbol\epsilon',\omega_{\bm{k}}) \tau_0 + \bm{T}(\boldsymbol\epsilon,\boldsymbol\epsilon',\omega_{\bm{k}}) \cdot \boldsymbol\tau \,,
\end{equation}
where $\tau_0$ is the $2\times2$ identity matrix, $\boldsymbol\tau=(\tau_{x},\tau_{y},\tau_{z})$ is the vector of Pauli matrices in (pseudo)spin space, and
\begin{align}
 T_0(\boldsymbol\epsilon,\boldsymbol\epsilon',\omega_{\bm{k}}) =&{\ } R_{a_{1g}^B}^{(0)}(\omega_{\bm{k}}) (\epsilon_x {\epsilon_x'}^* + \epsilon_y {\epsilon_y'}^*) + R_{a_{1g}^A}^{(0)}(\omega_{\bm{k}}) \epsilon_z {\epsilon_z'}^* \,,\\
 T_1(\boldsymbol\epsilon,\boldsymbol\epsilon',\omega_{\bm{k}}) =&{\ } R_{e_u}^{(1)}(\omega_{\bm{k}}) (\boldsymbol\epsilon \times \boldsymbol\epsilon') \cdot \hat{\bm{x}} \,,\\
 T_2(\boldsymbol\epsilon,\boldsymbol\epsilon',\omega_{\bm{k}}) =&{\ } R_{e_u}^{(1)}(\omega_{\bm{k}}) (\boldsymbol\epsilon \times \boldsymbol\epsilon') \cdot \hat{\bm{y}} \,,\\
 T_3(\boldsymbol\epsilon,\boldsymbol\epsilon',\omega_{\bm{k}}) =&{\ } R_{a_{2u}}^{(1)}(\omega_{\bm{k}}) (\boldsymbol\epsilon \times \boldsymbol\epsilon') \cdot \hat{\bm{z}} \,,
\end{align}\label{eq:dec}%
\end{subequations}
with $\hat{\bm{x}}, \hat{\bm{y}}, \hat{\bm{z}}$ unit vectors. The fundamental x-ray scattering spectra $R_{a_{1g}^B}^{(0)}, R_{a_{1g}^A}^{(0)}, R_{e_u}^{(1)}$ and $R_{a_{2u}}^{(1)}$ of the irradiated ions can be accurately determined via crystal-field calculations, as discussed in Ref.~\cite{Haverkort2010b}. In the above equations, it is seen that appropriate choice of cross-polarized x-rays can probe each of the spin components individually. For example, selecting $\boldsymbol\epsilon \parallel \hat{\bm{x}}$ and $\boldsymbol\epsilon' \parallel \hat{\bm{y}}$ yields
\begin{equation}
 {\cal F}_{fg}(\boldsymbol\epsilon \parallel \hat{\bm{x}} ,\boldsymbol\epsilon' \parallel \hat{\bm{y}},\bm{q},\omega_{\bm{k}}) \propto \braket{f| {\cal S}_{\bm{q}}^z |g} \,,
\end{equation}
where we have defined the quantity
\begin{equation}
 \bm{{\cal S}}_{\bm{q}} = \frac12 \sum_{\bm{k},s,s'} c_{\bm{k}+\bm{q},s}^\dagger \boldsymbol\tau_{s,s'} c_{\bm{k},s'} \,.
\end{equation}
All components of $\bm{{\cal S}}_{\bm{q}}$ can be obtained in this manner. This selectivity is still possible even when all the terms up to third order in spin are included (see Eq.~(16) of Ref.~\cite{Haverkort2010}) and regardless of the precise value of the fundamental x-ray scattering spectra, as long as these are nonzero.

In WSMs, the interesting physics happens close to the nodal points of the band structure. There, the effective electron spinor  has a pseudospin $\sigma=$``$\uparrow$'',``$\downarrow$'', with ``$\uparrow$'' and ``$\downarrow$'' signifying two orthogonal linear combinations of the original spin and orbital degrees of freedom of the material. This combination is material-dependent and can be accurately calculated within DFT. In this low-energy subspace, the excitation and decay operators are reduced to this pseudospin subspace: $c_{\bm{k}+\bm{q},s}^\dagger \rightarrow c_{\bm{k}+\bm{q},\sigma}^\dagger$ and $c_{\bm{k},s} \rightarrow c_{\bm{k},\sigma}$. The matrix $T$ is similarly downfolded to the pseudospin basis and a decomposition similar to that of Eq.~\ref{eq:dec} is obtained. In the special case where the orbital degree of freedom does not vary appreciably around the Weyl node, then the description of the winding is precisely that of Eq.~\ref{eq:dec}, i.e., $s\equiv\sigma$. For the case of TaAs, studied below, the pseudospin is actually a ``pseudo-orbital'' degree of freedom, arising from the $5d$ orbitals of Ta.

The respective RIXS cross sections then reduce to
\begin{subequations}
\begin{align}
 I_i(\bm{q},\omega_{\bm{kk}'}) = \sum_{\bm{k},b',b} &{\ } | F_{b'b}^i(\bm{q},\bm{k}) |^2 \nonumber\\ 
 &{\ } \times \delta(\varepsilon_{\bm{k},b} - \varepsilon_{\bm{k}+\bm{q},b'} + \hbar\omega_{\bm{kk}'}) \,,
\end{align}
with
\begin{equation}
 F_{b'b}^i(\bm{k},\bm{q}) = \braket{\psi_{\bm{k}+\bm{q},b'} | \tau_i | \psi_{\bm{k},b}} \,,
\end{equation}\label{eq:rixsxy}%
\end{subequations}%
where $i=x,y,z$ and $\ket{\psi_{\bm{k},b}}$ are the eigenstates of ${\cal H}$ at momentum $\bm{k}$ with band index $b=\pm$. For scattering from a partially filled lower band, $b=-$ and
\begin{subequations}
\begin{align}
 | F_{\pm-}^x |^2 =&{\ } \frac12 - \frac{g_{3,\bm{k}+\bm{q}} g_{3,\bm{k}}}{2g_{\bm{k}+\bm{q}} g_{\bm{k}}} \mp \frac{g_{\perp,\bm{k}+\bm{q}} g_{\perp,\bm{k}}}{2g_{\bm{k}+\bm{q}} g_{\bm{k}}} \cos(\phi_{\bm{k}+\bm{q}}+\phi_{\bm{k}}) \,,\\
 | F_{\pm-}^y |^2 =&{\ } \frac12 - \frac{g_{3,\bm{k}+\bm{q}} g_{3,\bm{k}}}{2g_{\bm{k}+\bm{q}} g_{\bm{k}}} \pm \frac{g_{\perp,\bm{k}+\bm{q}} g_{\perp,\bm{k}}}{2g_{\bm{k}+\bm{q}} g_{\bm{k}}} \cos(\phi_{\bm{k}+\bm{q}}+\phi_{\bm{k}})  \,,
\end{align}
\end{subequations}
where $g_{\perp,\bm{k}} = \sqrt{g_{1,\bm{k}}^2 + g_{2,\bm{k}}^2}$ and $\phi_{\bm{k}}=\mathrm{Arg}(g_{1,\bm{k}}+\mathrm{i} g_{2,\bm{k}})$. Fully analogous results are retrieved when $b=+$.

In crystals with low symmetry, a single measurement with cross-polarized beams may not be sufficient to isolate the pseudospin components. It is, however, always possible to obtain RIXS spectra for two or more different polarizations, which can then be appropriately added or subtracted to isolate the desired pseudospin density components themselves. For example, in the cuprates, where the relevant degree of freedom is simply the $3d_{x^2-y^2}$ orbital, two differently polarized measurements are enough to extract useful information about the orbital structure with RIXS~\cite{Marra2012}. The correct polarization combinations for the measurement of the topological charge can be deduced from a quantitative estimate of the participation of different orbitals at each $\bm{k}$-point in the band structure, which is readily achievable with density-functional theory (DFT). Since such detailed \textit{ab initio} calculations of the orbital structure around Weyl nodes of materials are lacking, in this proof-of-principle study we focus directly on the fundamental RIXS scattering amplitude of Eq.~\eqref{eq:rixsdef}. However, it should be stressed that accurate DFT calculations can readily provide all the necessary information for an experimental case study.

\section{Measurement of the topological invariant of Weyl nodes}\label{sec:meas}

We now define the difference
\begin{align}
 {\cal D}_{\pm-}(\bm{q},\bm{k}) =&{\ } |F_{\pm-}^y(\bm{q},\bm{k})|^2 - |F_{\pm-}^x(\bm{q},\bm{k})|^2 \nonumber\\
  =&{\ } \pm  \frac{g_{\perp,\bm{k}+\bm{q}} g_{\perp,\bm{k}}}{g_{\bm{k}+\bm{q}} g_{\bm{k}}} \cos(\phi_{\bm{k}+\bm{q}}+\phi_{\bm{k}}) \,,\label{eq:cos}
\end{align}
When $g_{\perp,\bm{k}+\bm{q}} g_{\perp,\bm{k}} \not= 0$, the zero-crossings of ${\cal D}_{\pm-}$ are given solely by the angle $\phi_{\bm{k}+\bm{q}}+\phi_{\bm{k}}$. The condition ${\cal D}_{\pm-}=0$ can be expressed equivalently as
\begin{equation}
 g_{1,\bm{k}+\bm{q}} g_{1,\bm{k}} - g_{2,\bm{k}+\bm{q}} g_{2,\bm{k}} = 0 \,.\label{eq:condition}
\end{equation}
Using the combined RIXS spectra, one can obtain the difference spectrum
\begin{align}
 I_{\cal D} (\bm{q},\omega_{\bm{k}},\omega_{\bm{k}'}) = \sum_{\bm{k},b'} &{\ } {\cal D}_{b'-}(\bm{q},\bm{k},\omega_{\bm{k}}) \nonumber\\ 
 &{\ } \times \delta(\varepsilon_{\bm{k},-} - \varepsilon_{\bm{k}+\bm{q},b'} + \hbar\omega_{\bm{kk}'}) \,,\label{eq:rixsdiff}
\end{align}
which is an experimentally attainable quantity.

\begin{figure}[t]
 \centering
 \includegraphics[width=\columnwidth]{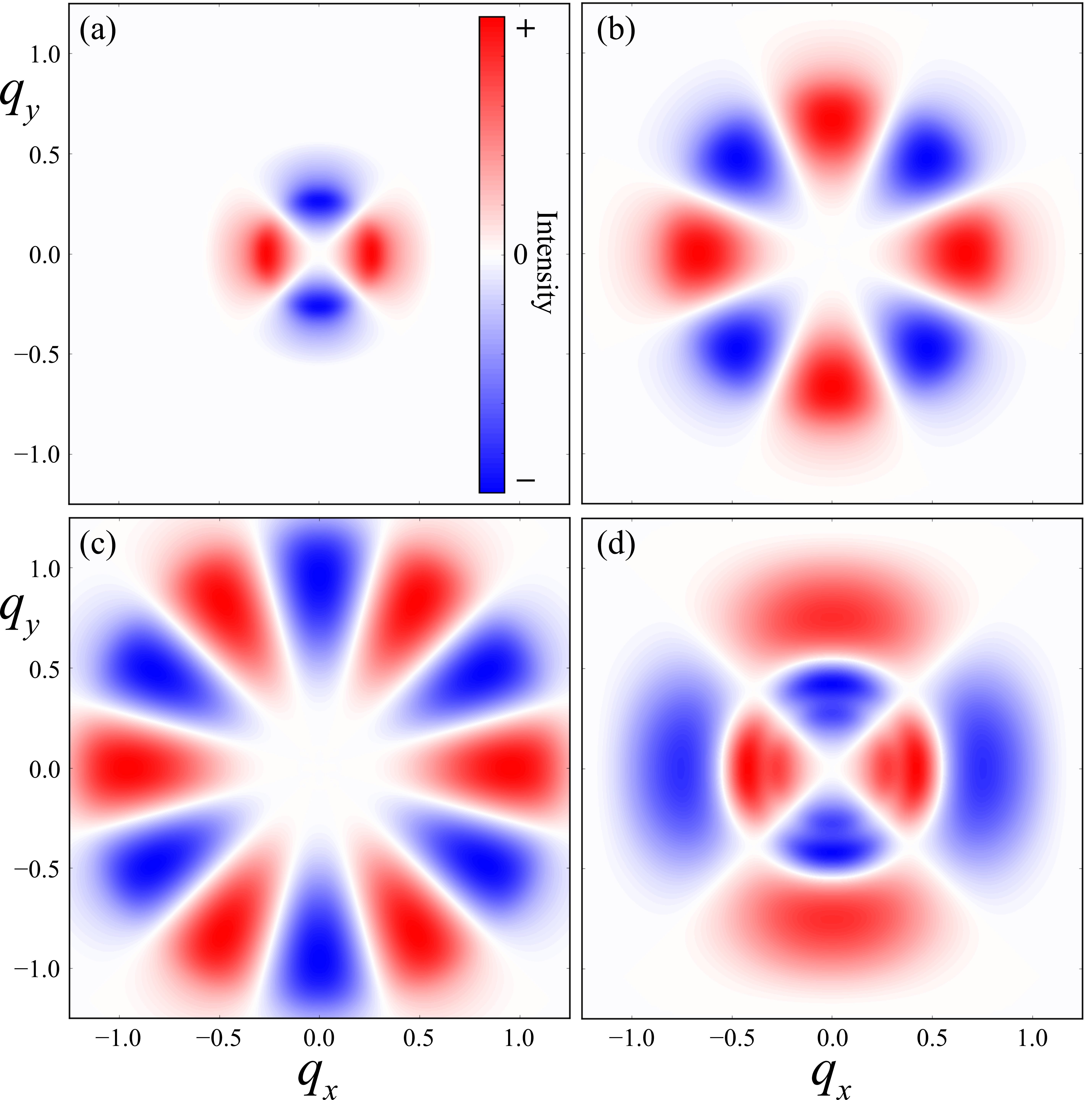}
 \caption{RIXS difference spectrum $I_{\cal D}$ at $\hbar \omega_{\bm{kk}'} = 0.25$ for (a) single Weyl node, (b) double Weyl node, (c) triple Weyl node, defined as $\bm{g}_{\bm{k}}^{\mathrm{triple}} = ( v_x k_x (v_x^2 k_x^2 - 3 v_y^2 k_y^2), v_y k_y (v_y^2 k_y^2 - 3 v_x^2 k_x^2), v_z k_z )$~\cite{Fang2012a} and (d) type-II single Weyl node with $g_{0,\bm{k}}=1.2k_z$. In all cases, $v_x=v_y=v_z=1$.}
 \label{fig:rixswnodes}
\end{figure}

The zeros of $I_{\cal D}$ reflect the winding of $\bm{g}_{\bm{k}}$ around the Weyl node. The condition of Eq.~\eqref{eq:condition} has the solutions $q_y = \pm q_x$ for a single Weyl node. There are therefore 4 solutions for $\varphi_{\bm{q}}=\mathrm{Arg}(q_x+\mathrm{i}q_y)\in[0,2\pi)$. For a double Weyl node, the solutions in the first quadrant are $q_y = \tan( \frac{\pi}{4} \pm \frac{\pi}{8} ) q_x$, and $I_{\cal D}$ becomes zero for 8 values of $\varphi_{\bm{q}}\in[0,2\pi)$ in total. Similarly, the RIXS difference spectrum for a triple Weyl node has 12 zero-crossings for $\varphi_{\bm{q}}\in[0,2\pi)$. Numerical calculation confirms that these are indeed the only solutions of Eq.~\eqref{eq:condition}. We therefore see that the number of zero-crossings of $I_{\cal D}$ for scattering around a Weyl node is $4|C|$, where $C$ is the topological charge. Note that the zero-crossings are the same regardless of whether the scattering is intra- or inter-band, as the band index does not enter the cosine in Eq.~\eqref{eq:rixsdiff}. This means that the measurement is possible in both elastic and inelastic scattering. This is important because, unlike quantum oscillations~\cite{Huang2015,Zhang2015,Zhang2015a,Klotz2015,Zhang2016,Arnold2016}, the RIXS measurement works even in the case where the Fermi surface that encloses a Weyl node is vanishingly small. The RIXS difference spectra $I_{\cal D}$ for inter-band scattering, calculated for a few idealized cases, is shown Fig.~\ref{fig:rixswnodes}.

In experiments on WSMs, the RIXS spectrum at small $\bm{q}$ will contain a superposition of contributions from at least two --- and commonly several --- Weyl nodes, as well as trivial Fermi surfaces. Even though trivial Fermi surfaces do not contribute a net Berry flux, the contributions from Weyl nodes will need to be disentangled in the resulting spectra, in order to extract the Chern numbers. Inter-node scattering between time-reversal or inversion symmetry partners, on the other hand, can come uniquely from a single pair of Weyl nodes, avoiding spurious contributions. This facilitates the targeted measurement of the topological charge of specific Weyl nodes in real materials. In the simple case of a linear spectrum around two Weyl nodes related by time-reversal symmetry, intra- and inter-node scattering are equivalent, as we show in the next Section.

\section{Geometry of RIXS around Weyl nodes and robustness against finite perturbations}\label{sec:geom}

\begin{figure}[t]
 \centering
 \includegraphics[width=\columnwidth]{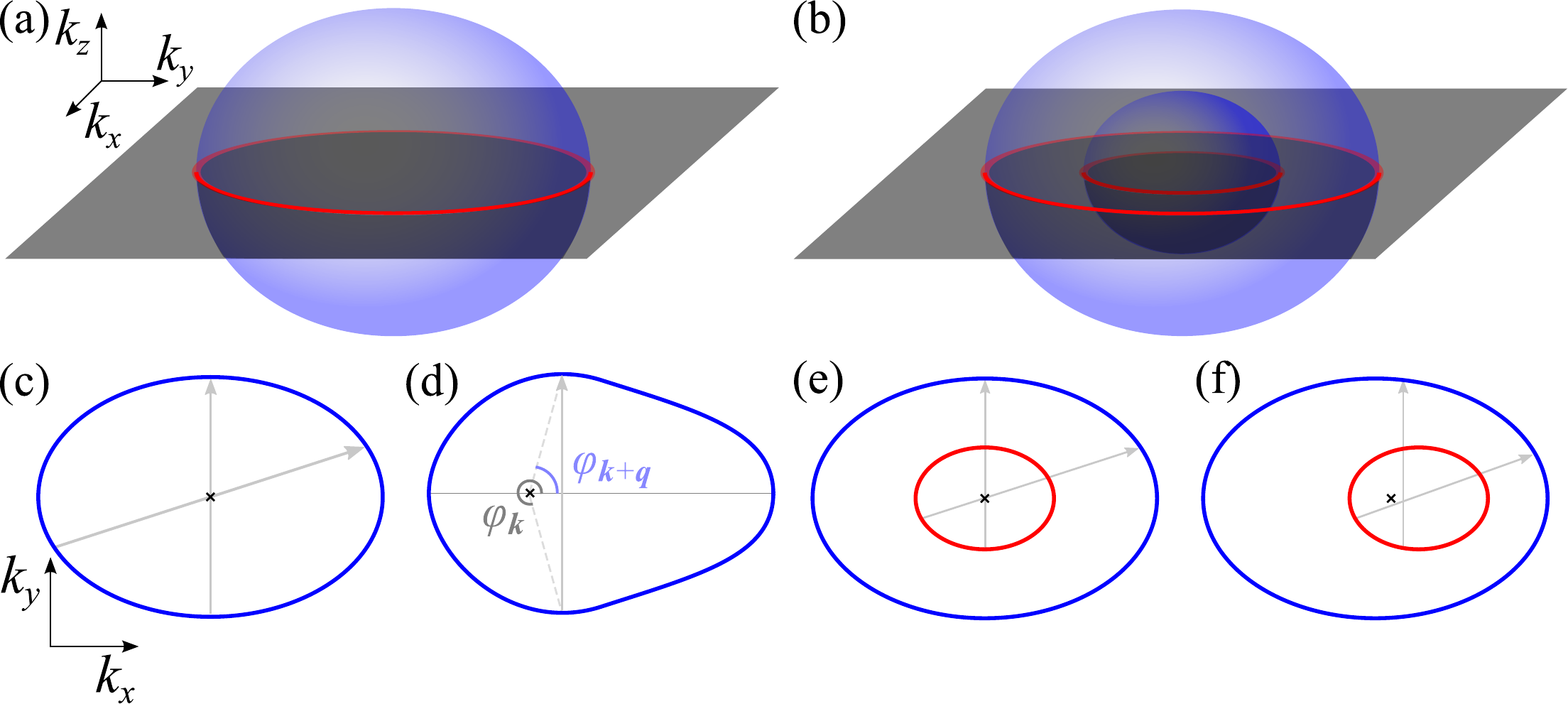}
 \caption{Cartoons of: (a) Fermi surface $\Omega$ and (b) constant-energy surfaces $\Omega$ and $\Omega'$, relevant to elastic and inelastic scattering around a Weyl node, respectively, (c,d) Fermi-surface contour in the $k_z = 0$ plane (c) for an unperturbed and (d) for a perturbed Weyl node hamiltonian, both with $\mu\not=0$, (e,f) constant-energy contours in the $k_z = 0$ plane (e) for an unperturbed and (f) for a perturbed Weyl node hamiltonian. In (d) the perturbation is such that the shape of the Fermi-surface contour is no longer elliptic, whereas in (f), even though the constant energy contours are still elliptic, the scattering is no longer from $\phi_{\bm{k}}$ to $\phi_{\bm{k}}+\pi$.}
 \label{fig:scat}
\end{figure}

Before turning to the geometry of RIXS scattering, consider the gauge-invariant quantity
\begin{equation}
 \Delta S_{\bm{k}} = | S_{\bm{k},\pm}^y |^2 - | S_{\bm{k},\pm}^x |^2 \,,
\end{equation}
where $S_{\bm{k},\pm}^i = \braket{\psi_{\bm{k},\pm}| \tau_i |\psi_{\bm{k},\pm}}$, $i=x,y$. Let ${\cal P}$ be a contour on the Fermi surface defined by $g_{\bm{k}} = \mu > 0$, with $g_{0,\bm{k}}=0$. In this case,
\begin{equation}
\Delta S_{\bm{k}} = - \frac14 \left( 1 - \frac{v_z^2 k_z^2}{\mu^2} \right)^{1/|C|} \cos 2 \phi_{\bm{k}} \,.
\end{equation}
When $v_z k_z \not= \pm \mu$, $\Delta S_{\bm{k}}$ changes sign and crosses zero at exactly 4 (8) angles $\phi_{\bm{k}}\in[0,2\pi)$ for single (double) Weyl nodes. We denote the number of zero-crossings of $\Delta S_{\bm{k}}$ on a path ${\cal P}$ around a Weyl node as $N$ and conclude that $N=4|C|$. Since the winding of the spin around the Weyl node is a property of topological origin, it is robust against any perturbation that leaves the topological charge of the Weyl node intact.

RIXS close to a Weyl node can be visualized geometrically. In the simplest case, $\varepsilon_{\bm{k},b} = \varepsilon_{-\bm{k},b}$ around the Weyl point. We shall consider the hamiltonian $\bm{g}_{\bm{k}}=\bm{g}_{\bm{k}}^{\mathrm{single}}$ and $g_{0,\bm{k}}=0$ for illustration, but everything we discuss below carries over straightforwardly to double or triple Weyl nodes. For a fixed energy transfer $\hbar\omega_{\bm{kk}'} = \hbar \omega_{\bm{kk}'} \geq 0$, scattering close to the node can take place between states on two concentric constant-energy surfaces $\Omega$ and $\Omega'$ in reciprocal space, for which the energy condition $\varepsilon_{\Omega',b'} - \varepsilon_{\Omega,b} = \hbar \omega_{\bm{kk}'}$ is fulfilled [see Fig.~\ref{fig:scat}(b)]. Of course, scattering takes place only from an occupied state on $\varepsilon_{\Omega,b}$ and an unoccupied one on $\varepsilon_{\Omega',b'}$. Assume a partially filled lower band and $\hbar \omega_{\bm{kk}'} = 0$, so that $b'=b=-$ and $\Omega \equiv \Omega'$ is simply the Fermi surface given by $g_{\bm{k}}=\mu$, as shown in Fig.~\ref{fig:scat}(a). Also, suppose that we choose to measure only scattering with $q_z=0$, so that scattering takes place within the constant-$k_z$ planes that intersect $\Omega$. Let us first treat elastic scattering. Turning summations into integrals and enforcing $\hbar \omega_{\bm{kk}'} = 0$, we obtain
\begin{align}
 I_{\cal D} (\bm{q},\omega_{\bm{k}}=\omega_{\bm{k}'}) \propto &{\ } \iint \mathrm{d}k_x \mathrm{d}k_y \cos(\phi_{\bm{k}+\bm{q}}+\phi_{\bm{k}}) \nonumber\\
 &{\ } \times \int \mathrm{d}k_z \left( 1-\frac{g_{3,\bm{k}}^2	}{\mu^2} \right) \,.\label{eq:rixsint}
\end{align}
For scattering at finite $q_x$ or $q_y$, the integral over $k_z$ in the second line above is always a positive factor. The largest scattering wavevectors in each constant-$k_z$ plane are those that correspond to scattering from $(-k_x,-k_y)$ to $(k_x,k_y)$, i.e., between antipodal points of $\Omega$. These are the wavevectors for which the condition $\cos(\phi_{\bm{k}+\bm{q}}+\phi_{\bm{k}}) = - \cos{2 \phi_{\bm{k}}}$ holds [$\cos(\phi_{\bm{k}+\bm{q}}+\phi_{\bm{k}}) = \cos{2 \phi_{\bm{k}}}$ for a double Weyl node]. The overall maximal scattering wavevectors occur in the equatorial scattering plane $k_z = 0$ and we denote them by $\bm{q}_{\phi_{\bm{k}}}$. For these wavevectors, the RIXS difference spectrum reduces to
\begin{equation}
 I_{\cal D} (\bm{q}_{\phi_{\bm{k}}},\omega_{\bm{k}}=\omega_{\bm{k}'}) \propto \cos{2 \phi_{\bm{k}}} \,,\label{eq:outerrim}
\end{equation}
since only scattering across the Fermi surface can take place. We thus recover the zero-crossings of $\Delta S_{\bm{k}}$ and therefore the topological charge of the Weyl node. These zero-crossings of $I_{\cal D}$ are exactly the ones that we found by explicit solution of Eq.~\eqref{eq:condition}. We note that our approach here bears some resemblance to that of Ref.~\cite{Li2015a}, although excitations in RIXS are in the particle-hole channel instead. In fact, for simple enough Fermi surfaces, the RIXS spectrum contains the Berry curvature seen by the particle-hole excitation in the outer shell of the three-dimensional scattering signal. This can then be integrated to yield the monopole charge, in full analogy to the paired-particle treatment of Ref.~\cite{Li2015a}, with the advantage that no pairing is required. However, unlike the paired case, the RIXS spectrum also contains all possible noncentrosymmetric contributions.

The zero-crossings of $I_{\cal D}$ are recovered in inelastic spectra as well. In this case, $\Omega$ and $\Omega'$ are distinct surfaces fulfilling the condition $\hbar \omega_{\bm{kk}'} > 0$. For $\varepsilon_{\bm{k},b} = \varepsilon_{-\bm{k},b}$ around the Weyl point, the largest scattering wavevectors within constant-$k_z$ planes correspond to transitions from $(\sqrt{k_x^2+k_y^2},\phi_{\bm{k}})$ to $(\sqrt{{k_x'}^2+{k_y'}^2},\phi_{\bm{k}}+\pi)$, with the overall largest wavevectors occuring for $k_z=0$. This again leads to the sinusoidal modulation of the intensity on the outer rim of the measured signal, as outlined above. This analysis carries over to double and triple Weyl nodes. In these cases, the shapes of the surfaces $\Omega$ and $\Omega'$ are more complicated, but the antipodal scattering condition in constant-$k_z$ planes [Eq.~\eqref{eq:outerrim}] holds as long as the condition $\varepsilon_{\bm{k},b} = \varepsilon_{-\bm{k},b}$ around the Weyl point is preserved.

We now discuss the effect of breaking the condition $\varepsilon_{\bm{k},b} = \varepsilon_{-\bm{k},b}$ around the Weyl node. The simplest perturbation that does that is $g_{0,\bm{k}} = \alpha k_i$, $i=x,y,z$, with $\alpha$ a real parameter. A finite $g_{0,\bm{k}}$ of this form does not affect the wavefunction; it only deforms the cone, while maintaining the linearity of the bands. This causes a relative shift between $\Omega$ and $\Omega'$ along one of the semi-principal axes. When this is the case, measurement of scattering in the plane perpendicular to the shift will still yield the topological invariant. Suppose, for example, that $g_{0,\bm{k}} = \alpha k_z$. Then, for RIXS measurements at $q_z = 0$, the scattering takes place between concentric constant-energy ellipses as before and we recover the same zero-crossings as above. This insensitivity allows one to obtain the topological charge of type-II Weyl nodes with RIXS in the same manner [see Fig.~\ref{fig:rixswnodes}(d)].

\begin{figure}[t]
 \centering
 \includegraphics[width=0.65\columnwidth]{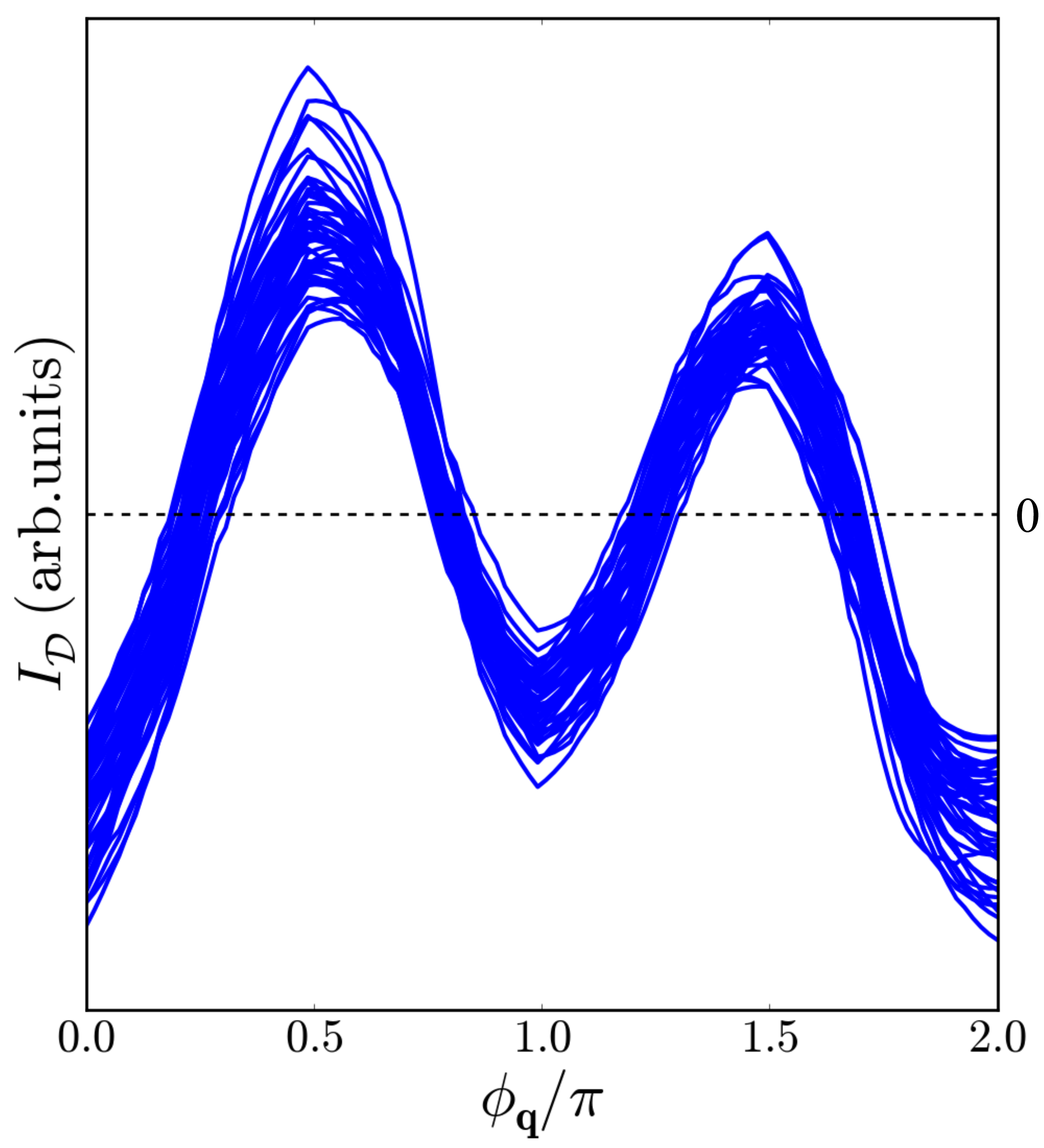}
 \caption{RIXS difference spectra $I_{\cal D}$ as a function of the polar coordinate $\phi_{\bm{q}}=\mathrm{Arg}(q_x + \mathrm{i}q_y)$ and $\sqrt{q_x^2+q_y^2}=0.5$ for 50 different realizations of a single Weyl node hamiltonian with $v_x=v_y=v_z=1$ and all possible perturbations of up to cubic order, each with a random prefactor in the range $[-0.2,0.2]$. All spectra cross zero (dashed line) at precisely 4 values of $\phi_{\bm{q}} \in [0,2\pi)$, confirming that finite perturbations do not affect the measurement of the topological charge of a Weyl node.}
 \label{fig:randweyl}
\end{figure}

To approximate a more realistic situation, we allow arbitrary perturbations of the simple Weyl-node hamiltonians we have treated so far. On one hand, topological properties, such as the zeros of $\Delta S_{\bm{k}}$, are robust against any perturbation that does not annihilate the Weyl node. On the other hand, perturbations that affect the scattering conditions, such as alterations of the shapes of $\Omega$ and $\Omega'$, can cause the features of $I_{\cal D}$ to deviate from those of $\Delta S_{\bm{k}}$. In particular, it is important to see whether perturbations can cause finite spectral weight to appear at the zero-crossings of $I_{\cal D}$. We argue that small perturbations can shift zero-crossings at the maximal scattering wavevectors $\bm{q}_{\phi_{\bm{k}}}$ but cannot remove them. Let us again consider only scattering for which $g_{\perp,\bm{k}+\bm{q}} g_{\perp,\bm{k}} \not= 0$ and concern ourselves only with the trigonometric factor $\cos(\phi_{\bm{k}+\bm{q}}+\phi_{\bm{k}})$, which defines the zero-crossings. When small perturbations are added to a Weyl-node hamiltonian, the shapes of the constant-energy surfaces $\Omega$ and $\Omega'$ are not exact ellipsoids anymore. Nevertheless, one may still define the overall largest scattering wavevectors $\bm{q}_{\phi_{\bm{k}}}$ that connect $\Omega$ and $\Omega'$. For small perturbations of the hamiltonian, the scattering conditions or both, we have
\begin{equation}
 I_{\cal D} (\bm{q}_{\phi_{\bm{k}}},\omega_{\bm{k}},\omega_{\bm{k}'}) \propto \cos(2 \phi_{\bm{k}} + \delta\phi_{\bm{k}}) \,,
\end{equation}
where we have defined $\phi_{\bm{k}+\bm{q}} \simeq \phi_{\bm{k}} + \pi + \delta\phi_{\bm{k}}$. The shift $\delta\phi_{\bm{k}} \ll \pi$ represents the perturbation to the geometry of the Fermi surface [see Figs.~\ref{fig:scat}(d,f) for schematic illustrations]. Since $\delta\phi_{\bm{k}}$ is small, we can expand the above as
\begin{align}
 \cos(2 \phi_{\bm{k}} + \delta\phi_{\bm{k}}) &= \cos(2\phi_{\bm{k}}) \cos \delta\phi_{\bm{k}} - \sin(2\phi_{\bm{k}}) \sin \delta\phi_{\bm{k}} \\
 &\approx \cos(2\phi_{\bm{k}}) - \delta\phi_{\bm{k}} \sin(2\phi_{\bm{k}}) \,.
\end{align}
This means that the perturbation can add a local shift of at most $\pm \delta\phi_{\bm{k}}$ to the sinusoidal modulation. This shift in turn moves the zero-crossings of $I_{\cal D}$, but as long as $\delta\phi_{\bm{k}}$ is small the zero-crossings cannot be removed. We have tested this robustness by adding random perturbations in the form of all possible terms of up to cubic order, each with a magnitude up to 0.2 times the linear velocity, to all components of $\bm{g}_{\bm{k}}$ and to $g_{0,\bm{k}}$ for single Weyl nodes and find that the signature of the topological charge remains unchanged (see Fig.~\ref{fig:randweyl}).

Even though the above discussion is concerned with intra-node scattering, the same principles carry over to inter-node scattering as well. In a time-reversal symmetric WSM, each Weyl node at $\bm{k}$ has a time-reversed partner at $-\bm{k}$ with the same chirality. For a linear spectrum, time reversal maps constant-energy contours around a Weyl node to identical contours around its partner. Inter-node scattering between time-reversal symmetry partners at momentum $\bm{Q}+\bm{q}$, with $\bm{Q}$ the separation between Weyl nodes, is therefore equivalent to intra-node scattering at momentum $\bm{q}$ in the linear regime close to the nodal points. Appropriate selection of the wavevector $\bm{Q}$ can yield scattering between only two Weyl nodes. We note that suitable $\bm{Q}$ vectors in a material can be deduced from density-functional theory (DFT), ARPES, or even RIXS itself whenever $\bm{Q}$ does not nest any other part of the Fermi surface apart from the two Weyl nodes of interest. As discussed above, small deformations of constant-energy contours cannot obscure the experimental signature of the topological charge of the nodes, as we exemplify in the case of TaAs below.

\section{Modeling of RIXS measurement in Tantalum monarsenide}\label{sec:taas}

\begin{figure}[t]
 \centering
 \includegraphics[width=\columnwidth]{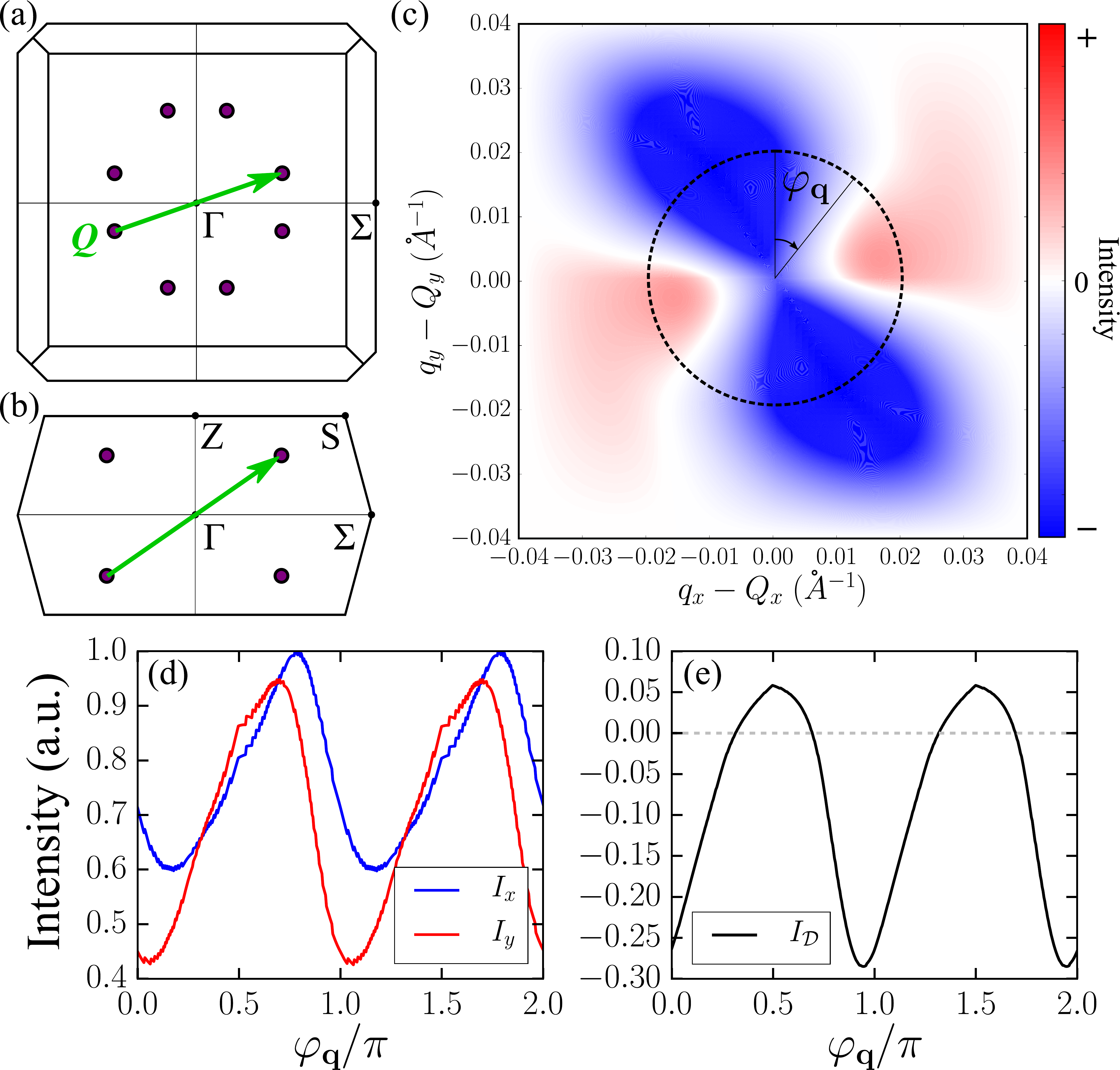}
 \caption{(a) Top and (b) side views of the BZ of TaAs, showing the projections of the 16 pairs of Weyl nodes. Two nodes project onto each purple dot. The distance between the $k_x$ and $k_y$ mirror-symmetric nodes is exaggerated for illustration. (c) RIXS difference spectrum $I_{\cal D}$, integrated over the range from 0 to 20~meV, for scattering between Weyl nodes connected by the wavevector $\bm{Q}$ shown in (a) and (b). (d) RIXS spectra $I_x$ and $I_y$, defined in Eq.~\eqref{eq:rixsxy}, calculated as a function of $\varphi_{\bf q}=\mathrm{Arg}[q_x - Q_x + \mathrm{i}(q_y-Q_y)]$ on the circle shown in (c), and (e) the corresponding difference spectrum $I_{\cal D}$. The intensities in (d) and (e) are normalized with respect to $\max(I_x,I_y)$. The small steps in (d) are due to the discretization of the circle in (c). The RIXS spectra are calculated for the $k\cdot p$ model derived from \textit{ab initio} calculations~\cite{Yu2015}, with the parameters used to fit experimental quantum oscillations data for TaAs~\cite{Huang2015} (see App.~\ref{app:taas})}
 \label{fig:taasw2}
\end{figure}

We exemplify inter-node scattering with an explicit calculation for TaAs. Of the 24 nodes in this material, 8 are in the $k_z = 0$ plane. Those at $k_z \not= 0$ are located in small Fermi-surface pockets, whose shape is roughly ellipsoidal~\cite{Lv2015b,Xu2015a,Zhang2016,Zhang2015a,Arnold2016}. Experiments reveal that the energy spectrum is to good approximation linear within $\sim20$~meV above and below the Fermi level~\cite{Lv2015b,Xu2015a,Yang2015}. The geometry of the Fermi surface can be accurately modeled by a linear $k \cdot p$ hamiltonian deduced from a full DFT calculation~\cite{Yu2015} that fits the Fermi surface determined by quantum oscillations in TaAs~\cite{Huang2015}. Using this hamiltonian, we evaluate the RIXS difference spectrum for scattering between one of the $k_z \not= 0$ Weyl nodes and its time-reversal partner. We calculate the wavefunction in a volume around each of the two Weyl nodes connected by the in-plane wavevector $\bm{Q}$ and evaluate the RIXS difference spectrum $I_{{\cal D}}$ for wavevectors $\bm{q}$ in the $q_x$-$q_y$ plane. The result is shown in Fig.~\ref{fig:taasw2}(b). The topological feature we anticipate, i.e., $4$ zero-crossings of $I_{{\cal D}}$ around $\bm{q}=\bm{Q}$, is clearly visible [see also Fig.~\ref{fig:taasw2}(e)]. The zero-crossings are obscured only in a small region at the center of the spectrum due to the tilting of the ellipsoidal Fermi surface. In all other respects, the RIXS difference spectrum is equivalent to that of the idealized single Weyl node shown in Fig.~\ref{fig:rixswnodes}. Both the energy and the momentum resolution required for this measurement are experimentally achievable today, as we discuss in Sec.~\ref{sec:exp}.

\begin{figure}[t]
 \centering
 \includegraphics[width=\columnwidth]{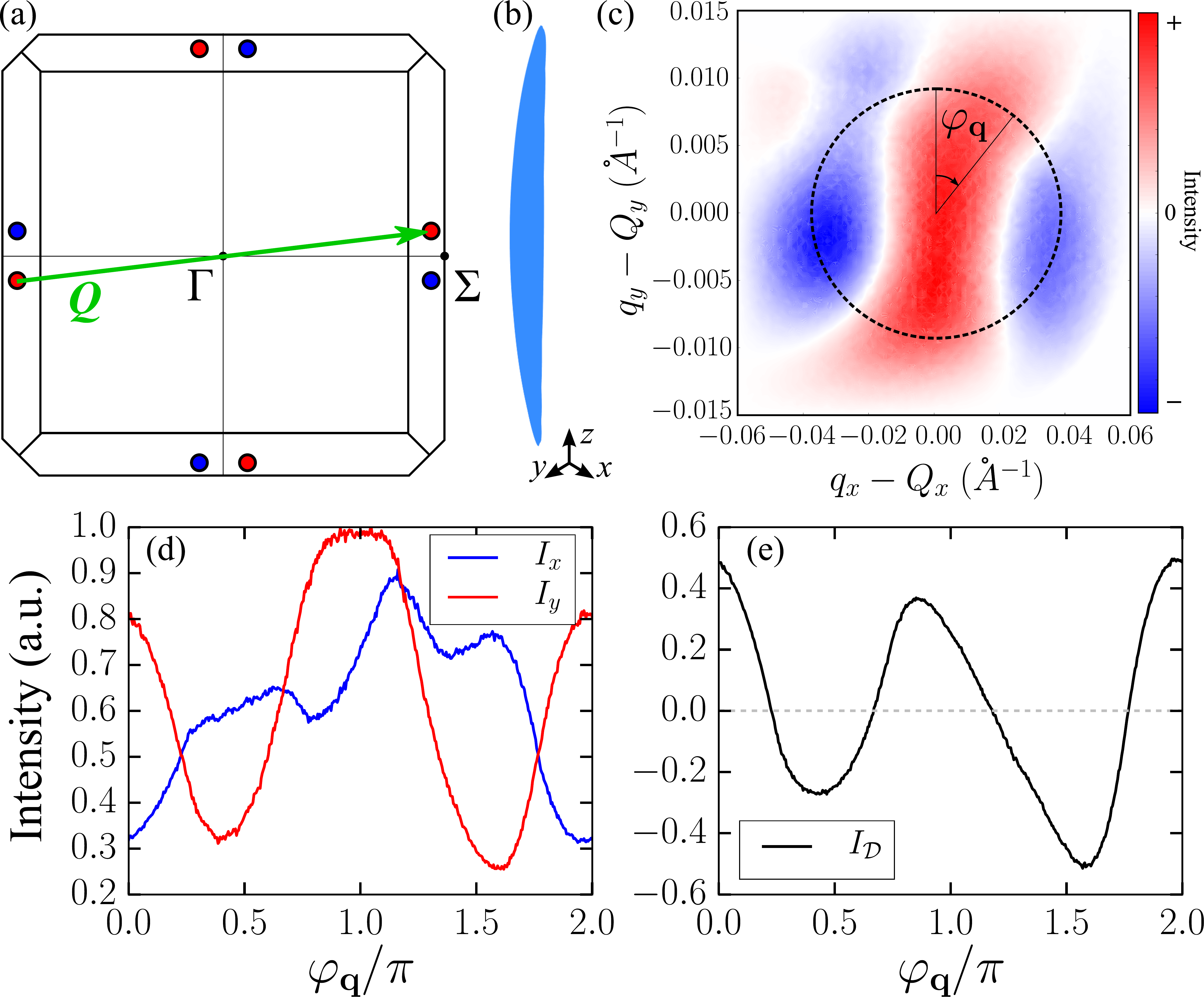}
 \caption{(a) $k_z = 0$ plane of the BZ of TaAs, showing the 4 pairs of Weyl nodes. The distance between the nodes close to each $\Sigma$ point is exaggerated for illustration. Scattering at the wavevector $\bm{Q}$ selects only two equivalent Weyl nodes, whereas scattering at $\bm{Q}'$ probes two pairs of nodes simultaneously. (b) Shape of the Fermi surface around one of the Weyl points; a more detailed visualization can be found in Ref.~\cite{Zhang2015a}. (c) RIXS difference spectrum $I_{\cal D}$, integrated over the range from 0 to 10~meV, for scattering between Weyl nodes connected by the wavevector $\bm{Q}$ shown in (a). (d) RIXS spectra $I_x$ and $I_y$, defined in Eq.~\eqref{eq:rixsxy}, calculated as a function of $\varphi_{\bf q}$ shown in (c), and (e) the corresponding difference spectrum $I_{\cal D}$. The intensities in (d) and (e) are normalized with respect to $\max(I_x,I_y)$. The RIXS spectra are calculated for the $k\cdot p$ model derived from \textit{ab initio} calculations~\cite{Weng2015}, with the parameters used to fit experimental ARPES data for TaAs~\cite{Zhang2016} (see App.~\ref{app:taas}). Note the unequal ranges in $q_x$ and $q_y$, reflecting the peculiar shape of the Fermi surface pocket shown in (b).}
 \label{fig:taasw1}
\end{figure}

The equivalence of intra- and inter-node scattering exploited above is an artifact of the linearity of the hamiltonian. We now demonstrate the topological charge measurement in inter-node scattering in the presence of quadratic and cubic terms by an explicit calculation for the Weyl nodes of TaAs in the $k_z = 0$ plane. The picture supported by experiment is that of a pair of Weyl nodes close to each of the four $\Sigma$ points of the BZ, located a few meV below the Fermi level~\cite{Lv2015b,Xu2015a}. The band structure close to one of these nodes can be accurately modeled using a $4\times4$ $k \cdot p$ model derived from the results of DFT calculations~\cite{Weng2015}. The correct parameters for TaAs were obtained by comparison to experimental magnetotransport and ARPES measurements~\cite{Zhang2016}. This model yields an oblong boomerang-shaped Fermi surface~\cite{Zhang2015a,Arnold2016}, indicating the sizable quadratic terms present in the hamiltonian. The electron spin varies very little close to Weyl nodes in both the DFT and the $k \cdot p$ model, so we can simply trace over the spin degree of freedom. The result is shown in Fig.~\ref{fig:taasw1}(c). Again we recover the $4$ zero-crossings of $I_{{\cal D}}$ around $\bm{q}=\bm{Q}$, with only a small region at the center of the spectrum obscured due to the irregular shape of constant-energy contours around the Weyl node. The robustness against nonlinear terms, which can be understood on the more general grounds discussed in Sec.~\ref{sec:geom}, shows that topological features can be recovered by RIXS even when band structures are quite complicated.

\section{Experimental feasibility}\label{sec:exp}

Our work shows how RIXS can be deployed to the study of topologically nontrivial band structures. Even though recent experimental efforts are directed towards using RIXS to understand properties of strongly correlated materials~\cite{Ament2011}, the technique has been very successful in inferring the band structure of approximately noninteracting systems~\cite{Johnson1994a,Carlisle1995,Jia1996,Carlisle1999,Denlinger2002,Kokko2003,Strocov2004a,Strocov2004,Ahn2009}. RIXS offers options that are not available to other techniques: spin and orbital selective scattering that allows for momentum-resolved determination of pseudospin components in the bulk, insensitivity to surface imperfections and arbitrarily large electromagnetic fields, direct access to the unoccupied band structure that eliminates the need for a measurable Fermi surface. This potential is especially tantalizing in settings where other spectroscopies are difficult to employ: the possibility for imaging of Dirac or WSM band structures in arbitrarily strong electromagnetic fields means that RIXS offers a unique platform for the direct spectroscopic observation of elusive physical phenomena, such as the chiral anomaly~\cite{Nielsen1983,Aji2012,Zyuzin2012a,Hosur2013a,Kim2013,Parameswaran2014}.

We estimate that already existing RIXS instrumentation is sufficient to perform the measurement proposed in this work. For real materials, the polarization dependence of the RIXS cross section can be intricate. This means that one may need to perform two or more measurements with different polarization combinations in order to isolate the desired signals. Even though cumbersome, this is certainly feasible: the polarization selectivity required has been successfully demonstrated experimentally at the $L_3$ edge of copper~\cite{Braicovich2014}. The energy resolution typically achieved in experiments is nominally in the desired range. For the case of TaAs (NbAs), the states close to the Fermi level are predominantly formed by the $5d$ ($4d$) electrons of Ta (Nb) (see, for example, the relevant calculation for TaAs in the Supplemental Material of Ref.~\cite{Inoue2016}). Considering that the Weyl nodes in TaAs were resolved with $\hbar \omega_{\bm{kk}'} \sim 50 - 80$~meV~\cite{Lv2015b,Zhang2016}, a similar resolution would be enough for the experiment we propose. The $L$ edges of Ta (Nb) are below 12~keV (3~keV). Existing beamlines can access this energy range using spectrometers with resolutions comparable to that of ARPES. Finally, both Ta and Nb $M$ and $N$ edges are in the soft x-ray regime ($\hbar\omega_{\bm{k}}<500$~eV) and are therefore even more easily accessible, while the optical elements currently in use in detectors can offer resolutions of the order of 10 meV, which is comparable to the best resolution of current ARPES spectra of TaAs~\cite{Yang2015}. When the separation between Weyl points in reciprocal space is small, high momentum resolution is necessary. Adequate resolution for the proposed measurement of the $k_z\not=0$ Weyl nodes is achievable in RIXS. As an example, we mention that energy and momentum resolutions of $\sim30$~meV and $\sim0.03~\mathrm{\AA}^{-1}$ have been reported for the $L_3$ edge of Ir already some years ago~\cite{Kim2011,Kim2014}, whereas modern synchrotron facilities improve upon these figures by a large margin~\cite{Horiuchi2015}. Finally, recent magnetotransport and quantum oscillation measurements show that Weyl nodes are present in TaIrTe${}_4$~\cite{Khim2016,Koepernik2016} and detailed DFT calculations predict that Mn${}_3$Sn and Mn${}_3$Ge are also WSMs~\cite{Yang2016}. This means that existing spectrometers designed for Ir and Mn ions can be used to probe Weyl nodes in these materials.

As an experimental technique, RIXS has particular complications associated with it, which may depend on the specifics of the material under investigation. Threshold singularities~\cite{Nozieres1969,Nozieres1974} constitute one such complication. For the measurement of Weyl nodes proposed here, this issue can be avoided simply by detuning away from resonance~\cite{Ahn2009}, since the measurement does not depend on satisfying the resonance condition exactly. The drawback of this detuning is that the photon count will drop, resulting in longer beamtimes. Another complication that may be of concern in some materials is deexcitation via the ``fluorescence'' channel, i.e., the case where an electron coming from a ``broad band'' fills the core hole. This may happen when a material is not so strongly correlated, or when it is a metal. This effect, however, is material specific and successful RIXS measurements of elementary excitations have been performed even in cases where fluorescence intensity is high~\cite{Zhou2013a}.

In RIXS, the surface of materials is mostly transparent and consequently surface states cannot be detected. On the other hand, complications related to surface preparation are bypassed. RIXS requires only small sample volumes and can access the entire BZ. It is insensitive to electromagnetic fields, a feature that may facilitate a direct spectroscopic detection of the chiral anomaly, even though the latter may also be accessible to high-resolution ARPES measurements~\cite{Behrends2015}. Finally, it should be noted that additional versatility may be afforded by careful examination of experimental RIXS data. For example, an analysis of time scales, i.e., inverse energy scales relevant to a material (hoppings, spin-orbit coupling), may offer more in-depth information on the individual processes that make up the full RIXS response~\cite{Bisogni2013,Wray2015}.

In the case where the pseudospin is indeed the real electron spin, the response defined in Sec.~\ref{sec:theory} reduces to the measurement of the dynamic spin structure factor with magnetic RIXS~\cite{deGroot1998,vanVeenendaal2006,Ament2009a}, which will contain the zero-crossings indicating the winding of the spin around the Weyl node. Inelastic neutron scattering (INS) could also be used to obtain the same magnetic signature, as it measures the same quantity. However, in existing WSMs, it is the orbital degree of freedom that winds around the Weyl point and not the spin. Furthermore, even in the case of a material with spin winding, INS would require large single crystals which may be difficult to grow.

The methodology presented in this work can be straightforwardly extended to scattering between inequivalent nodes, where the outcome may vary. As an inelastic probe, RIXS can also access features by scattering electrons across an energy gap. It is therefore conceivable that it can be used for the detection of bulk topological properties in gapped systems.

\begin{acknowledgments}
 The author thanks B.~A.~Bernevig for insightful comments and encouragement throughout all stages of this work. He is also grateful to A.~G.~Grushin, P.~Kotetes, T.~Neupert and J.~van den Brink for critical reading and discussion of the manuscript, L.~J.~P.~Ament, V.~Bisogni, C.~Mazzoli, N.~C.~Plumb and M.~Shi for invaluable input on RIXS modeling and experiments, and J.~Li and Z.~Wang for useful discussions of the band structures of WSMs. This work was supported financially by NSF CAREER DMR-0952428, ONR-N00014-11-1-0635, the Keck grant, and ICAM branch contributions.
\end{acknowledgments}

\appendix

\section{Modeling of the Weyl nodes in tantalum monarsenide}\label{app:taas}

To model the Weyl nodes in TaAs, we use two models derived via independent DFT calculations fitted to two different experimental measurements. The first is a linear $k \cdot p$ theory presented in Ref.~\cite{Yu2015}, which was designed to fit the DFT band structure close to the $k_z \not= 0$ Weyl nodes and match the experimentally observed Fermi surface size and shape~\cite{Huang2015}. The hamiltonian matrix is given by~\cite{Yu2015}
\begin{subequations}
\begin{equation}
H_{k_z \not= 0} = \mu \tau_0 + \sum_{i,j} k_{i} a_{ij} \tau_{j} \,,
\end{equation}
where $i,j=x,y,z$, $\tau_{j}$ and $\tau_0$ are the Pauli and identity matrices in pseudospin space and
\begin{equation}
a_{ij}=\left[\begin{array}{ccc}
2.657 & -2.526 & 0.926\\
0.393 & -2.134 & 3.980\\
-1.200 & -3.530 & 1.193
\end{array}\right] \, 
\end{equation}\label{eq:w2}%
\end{subequations}%
The precise position of the Fermi level does not affect the salient features of the RIXS spectrum. For the calculation in Sec.~\ref{sec:taas} we have chosen $\mu=2$~meV.

The pockets around the $k_z = 0$ Weyl nodes have a complicated shape that cannot be captured by linear terms alone. By including quadratic and cubic terms, the $k \cdot p$ introduced in Ref.~\cite{Weng2015} can be used to accurately model the band structure around the $k_z = 0$ Weyl nodes. This model has been used in Ref.~\cite{Zhang2016} to reproduce the ring-shaped trivial Fermi surface and the correct location and number of Weyl nodes obtained in DFT results, which were used to interpret ARPES and magnetotransport measurements. The hamiltonian matrix obtained after the fitting is
\begin{subequations}
\begin{align}
H_{k_z=0} =&{\ } H_0+H_{\mathrm{mass}} \,,\\
H_0=&{\ } e(\bm{k})\tau_0+d_1(\bm{k})\tau_x+d_2(\bm{k})\tau_y+d_3(\bm{k})\tau_z \,,\\
H_{\mathrm{mass}}=&{\ }
m_1(\bm{k})\tau_0s_y+m_2(\bm{k})\tau_zs_y+m_3(\bm{k})\tau_xs_x \\ 
&{\ }+\,m_4(\bm{k})\tau_xs_z+m_5(\bm{k})\tau_ys_z+m_6(\bm{k})\tau_ys_x \,,\nonumber
\end{align}
where $s_i$ ($\tau_i$), $i=0,x,y,z$ are the identity and Pauli matrices spanning the spin (effective orbital) degree of freedom, and
\begin{eqnarray}
e(\bm{k})&=&\mu+w k_x +\mathcal{O}(\bm{k}^2),\\
d_1(\bm{k})&=&uk_yk_z+\mathcal{O}(\bm{k}^3),\\
d_2(\bm{k})&=&vk_y+\mathcal{O}(\bm{k}^2),\\
d_3(\bm{k})&=&M-\alpha k_x^2- \beta k_z^2+ \gamma k_x
\nonumber\\
&&+ \zeta k_x^3+\mathcal{O}(\bm{k}^3),\\
m_1(\bm{k})&=&m_1 +\mathcal{O}(\bm{k}),\\
m_2(\bm{k})&=&m_2 +\mathcal{O}(\bm{k}),\\
m_3(\bm{k})&=&m_3 k_z+\mathcal{O}(\bm{k}^2),\\
m_4(\bm{k})&=&m_4 +m_4^\prime k_x +\mathcal{O}(\bm{k}^2),\\
m_5(\bm{k})&=&m_5 k_z+\mathcal{O}(\bm{k}^2),\\
m_6(\bm{k})&=&m_6 +\mathcal{O}(\bm{k}) \,,
\end{eqnarray}\label{eq:w1}%
\end{subequations}%
with $M=12.23$, $\mu=-3.504$, $u=-763.1$, $v=-685.1$, $w=34.11$, $\alpha=682.8$, $\beta=583.0$, $\gamma=264.2$, $\zeta=-147.5$, $m_1=7.019$, $m_2=1.031$, $m_3=0.9078$, $m_4=0.0$, $m_4^\prime=-11.07$, $m_5=-56.50$, $m_6=-4.097$, all in units of meV and the appropriate power of $\textrm{\AA}$. This model does not account for the correct position of the chemical potential, so we shift the energy spectrum by an additional $+4$~meV, in order to reproduce the size and shape of the Fermi surface predicted by DFT. Note that the prefactors $u, \alpha, \beta, \zeta$ of quadratic and cubic terms are large compared to those of linear terms, meaning that this model is away from the idealized linear limit.

\section{RIXS difference spectrum for generic nodes}

\begin{figure}[t]
 \centering
 \includegraphics[width=\columnwidth]{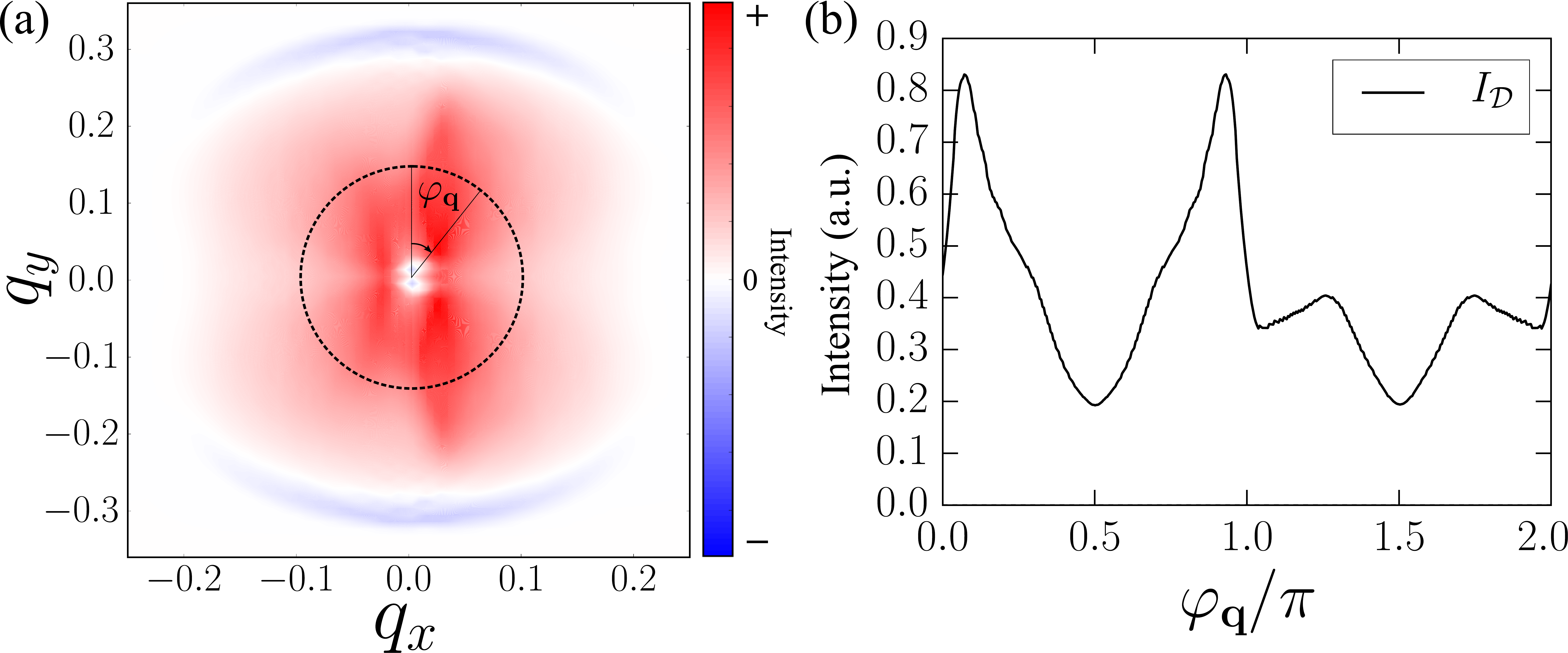}
 \caption{(a) $q_x$-$q_y$ intensity map and (b) spectrum as a function of the polar coordinate $\phi_{\bm{q}}=\mathrm{Arg}(q_x + \mathrm{i}q_y)$ of the RIXS difference $I_{\cal D}$, integrated over the range from 0 to 80~meV, for the trivial node in Cd${}_3$As${}_2$ with an applied magnetic field, as described by the low energy theory of Eq.~\eqref{eq:nonweyl}~\cite{Wang2013a,Cano2016}. The parameter values used are $M_1 = A = 1~\mathrm{eV\AA}$, $M_2 = A_y = 1~\mathrm{eV\AA}^2$, $B_1 = B_2 = 1~\mathrm{eV\AA}^3$ and $\mu = 0.4~\mathrm{eV}$.}
 \label{fig:nonweyl}
\end{figure}

In this section, we give an example of the methodology developed in this work, in which the node studied is topologically trivial, i.e., has zero monopole charge. Such nodes appear when there are higher-order crossing points in a band structure. For example, such a node can be found in the band structure of Cd${}_3$As${}_2$ in a magnetic field along [100]~\cite{Wang2013a,Cano2016}. The relevant low-energy theory for this node can be written simply as~\cite{Wang2013a,Cano2016}
\begin{subequations}
\begin{align}
 H_{C=0} =&{\ } -2 [ k_y M_1 + (k_x^2 + k_z^2) M_2 ] \tau_z \\
 &{\ } - 2(B_1 + B_2)(k_x^2 - k_z^2)k_y \tau_x \\
 &{\ } - A k_z (1 + A_y k_y^2) \tau_y + \mu \tau_0 \,,
\end{align}\label{eq:nonweyl}
\end{subequations}%
where $M_1$, $M_2$, $B_1$, $B_2$, $A$, $A_y$ and $\mu$ are all constants. 

The RIXS difference spectrum for intra-node scattering is shown in Fig.~\ref{fig:nonweyl}. There are no zero-crossings of $I_{\cal D}$, exactly as expected from the $4|C|$ rule derived in Sec.~\ref{sec:meas}. The same holds in any other $\bm{q}$-plane for this node. This example thus illustrates that topological and trivial nodes can be distinguished from their different responses to polarized resonant x-rays.

It is also interesting to briefly consider the RIXS response of a Dirac semimetal. In these materials, nodes are spin-degenerate due to time-reversal symmetry. The minimal low-energy description around a node requires a $4\times4$ Dirac spinor, which, in the simplest case, is just two antichiral copies of the Weyl equation. Although more complicated, a RIXS measurement similar to the one presented here may still be possible, exploiting the combined spin-orbital selectivity of the method. Despite the similarities between the two classes of materials, Dirac and Weyl semimetals are distinct and can even be distinguished by their RIXS responses in a magnetic field. Breaking time reversal in a Dirac semimetal leads to splitting of Dirac nodes. Since time-reversal symmetry breaking in the laboratory entails sizable magnetic fields, angle-resolved photoemission experiments cannot be used to capture this splitting. In these cases, RIXS is a viable method to infer the band structure around the split Dirac nodes, even without the need for polarization resolution, along the lines of Refs.~\cite{Johnson1994a,Carlisle1995,Jia1996,Carlisle1999,Denlinger2002,Kokko2003,Strocov2004a,Strocov2004,Ahn2009}. We defer a detailed discussion of this application of RIXS to a dedicated study.

\bibliographystyle{apsrev4-1}

\end{document}